\documentclass[a4paper,12pt]{article}
\usepackage{graphicx}
\usepackage{amsmath}
\usepackage{epstopdf}
\usepackage{hyperref}

\def\slashchar#1{\setbox0=\hbox{$#1$}           
   \dimen0=\wd0                                 
   \setbox1=\hbox{/} \dimen1=\wd1               
   \ifdim\dimen0>\dimen1                        
      \rlap{\hbox to \dimen0{\hfil/\hfil}}#1 
   \else                                        
      \rlap{\hbox to \dimen1{\hfil$#1$\hfil}}/                                    \fi}          
\def\notslashchar#1{\setbox0=\hbox{$#1$}           
   \dimen0=\wd0                                 
   \setbox1=\hbox{/} \dimen1=\wd1               
   \ifdim\dimen0>\dimen1                        
      \rlap{\hbox to \dimen0{\hfil\phantom{/}\hfil}}#1 
   \else                                        
      \rlap{\hbox to \dimen1{\hfil$#1$\hfil}}/                                    \fi}          

\newcommand\MTGEN{{$m_{T\mathrm{Gen}}$}}
\newcommand\MTTGEN{{$m_{TT\mathrm{Gen}}$}}
\newcommand\MTTWO{{$m_{T2}$}}
\newcommand\mtgen{{m_{T\mathrm{GEN}}}}
\newcommand\mttgen{{m_{TT\mathrm{GEN}}}}
\newcommand\pmiss{{{\slashchar{p}}}}
\newcommand\ptmiss{\slashchar{p}_T}

\def\ntlinoone{{\chi^0_1}}
\def\ntlinotwo{{\chi^0_2}}

\def\mttwosq{{m_{T2}^2}}

\def\half{{\frac 1 2}}

\def\gluino{{\tilde g}}
\def\squark{{\tilde q}}
\def\slepton{{\tilde l}}

\newcommand{\mysecref}[1]{section~\ref{#1}}

\newcommand{\myfigref}[1]{figure~\ref{#1}}
\newcommand{\definmath}[2] {\def#1{\ifmmode#2\else$#2$\fi}}
\definmath\amin{\mathrm{min}}
\definmath{\cht}{{\tilde{\chi}}}
\definmath{\DeltaMChi}{{\Delta M_{\cht_1}}}

\def\meff{{M_{\rm Eff}}}

\newcommand\mtTwo{m_{\rm T2}}
\newcommand\roots{{\sqrt s}}
\newcommand\what{{\hat w}}

\newcommand\quarter{{\frac{1}{4}}}
\newcommand\sigpt{{{\bf \underline \sigma}}}

\begin{document}

\vskip-15em\hskip-3em
  \underline{\small Cavendish-HEP-2007-05}
\\\vskip5mm
\begin{center}{\LARGE{\MTGEN\ : Mass scale measurements in
  pair-production at colliders
}\vspace{7mm}}
\begin{tabular}{ccc}
Christopher G.~Lester$^\dag$ 
and 
Alan J.~Barr$^\ddag$
\end{tabular}\\\vspace{3mm}
{\em $^\dag$ Cavendish Laboratory, J.J.Thomson Avenue, Cambridge,
        CB3\nolinebreak\ \nolinebreak{0HE,} UK}
\\
{\em $^\ddag$ Department of Physics, University of Oxford, 
              Keble Road, Oxford,
	      OX1\nolinebreak\ \nolinebreak{3RH,} UK}
\\\vspace{2mm}
Email address: $^\dag$lester@hep.phy.cam.ac.uk, 
$^\ddag$a.barr@physics.ox.ac.uk
\end{center}
\vspace{2mm}
\begin{center}
\begin{minipage}{.9\linewidth}
{\bf We introduce a new kinematic event variable \MTGEN\ which can
provide information relating to the mass scales of particles
pair-produced at hadronic and leptonic colliders. The variable is of
particular use in events with a large number of particles in the final state
when some of those particles are massive and not detected,
such as may arise in R-parity-conserving supersymmetry.}
\end{minipage}
\end{center}

\section{Introduction}

When LHC experiments
like ATLAS \cite{atlasphysicstdr1,atlasphysicstdr2} and CMS 
\cite{cmsphystdr}
begin to look for signs of supersymmetry (or any other models with
large numbers of new particles), one of the first things they will want
to probe is the mass scale associated with these new particles, if
there are any.  Single particles produced in narrow $s$-channel
resonances will in general be easy to spot (one would hope) and will
not be considered further here.  The fun begins when new particles are
produced in pairs in non-resonant processes, 
in particular those in which some particles go unobserved.
Examples within the context of R-parity conserving supersymmetry might 
include processes like $gg\rightarrow
\gluino\gluino$, $gq\rightarrow \gluino\squark$, $q\bar q \rightarrow
\ntlinotwo \ntlinoone$ or $q\bar q \rightarrow \slepton \bar
\slepton$.

\par

In this paper we present a new event variable, ``\MTGEN'', which is
designed to measure the mass scale(s) associated with any new
particle(s) which might be {\em pair} produced at future colliders. On
an event-by-event basis, \MTGEN\ supplies a lower bound for the mass
of either of the two particles which were pair produced and whose
decay products were observed, under the assumption that the event was
indeed of that kind.  The intention is that over {\em many}
events a histogram of \MTGEN\ would reveal one or more edge structures
whose upper endpoints would correspond to the masses of the particles
which were being produced in large numbers.  The \MTGEN\ variable
takes as input (1) the reconstructed momenta of the observed particles
in each event, and (2) the masses of any unobserved particles which
are taken to have been produced in the decays of the primary particles.
Though \MTGEN\ can benefit from information regarding particle
identification, this is not a requirement.

\par

It is not the intention of this paper to discuss how \MTGEN\ performs
when particle momenta are poorly measured, nor to discuss
issues of finite precision or acceptance.
Demonstration of the performance of \MTGEN\ in environments
representative of future colliders is the subject of current ongoing work.
This paper seeks primarily (1) to provide a source of documentation for the
definition of the \MTGEN\ variable, and (2) to document in the
Appendix an analytic closed form approximation for \MTTWO\ which is
valid for events with very little initial state radiation and which is
needed to demonstrate that \MTGEN\ can be calculated efficiently.

\section{\MTGEN}

As far as \MTGEN\ is concerned, an event at a future collider is a set
$O$ containing $n_O$ observed Lorentz four-momenta: $O = \{o_i^\mu :
i=1,\ldots, n_O\}$.  Although quantum interference means that terms
like ``initial state'' and ``final state'' cannot really be applied to
the momenta in $O$ in a well defined way, it is nevertheless common
and expedient when analysing real events to treat some momenta as if
they were the result of the decays from the pair of partons produced
in the primary $2\rightarrow2$ process used in the matrix element, and
to treat other momenta as if they were the result of initial state
radiation (ISR).  We will therefore divide the observed momenta into
two non-overlapping sets $F= \{f_i^\mu : i=1,\ldots, n_F\}$ (for
momenta supposedly from the central $2\rightarrow 2$ pair production
process) and $G=\{g_i^\mu : i=1,\ldots, n_G\}$ (for momenta supposedly
from initial state radiation).  In practice this might be done by
assigning all four-momenta whose transverse momenta are greater than
some threshold and whose rapidity is sufficiently central to $F$,
while placing anything else in $G$.  The value of \MTGEN\ for each
event will inevitably depend to some extent the exact nature of the
cuts used, and potential biases would have to be investigated in
specific cases.  Most event variables (e.g.\ thrust and sphericity)
have similar second order dependence on cuts.  The endpoint structures
which are the eventual target of any \MTGEN\ investigation, however,
are expected to be particularly insensitive to these cuts.  Individual
events where misassignments are made will either be swept below the
endpoint by the minimisation procedure (when momenta are omitted from
$F$), or will be smeared above the endpoint (when ISR with unusually
large transverse momentum is added to $F$ in error).

In a real event, we do not know from which ``side''\footnote{We will
use the term ``side'' to refer to the division of the particles in $F$
into two groups, depending on which of the two outgoing primary
particles they descend from.  An event, then, is an object
with two sides, and possibly also some initial state radiation.
The term ``side'' is not meant to suggest that the momenta of a
particular side are in some way spatially correlated (e.g.\ in one
hemisphere).  Indeed, if the two primary partons were scalars produced
at threshold, then the decay products of each ``side'' would be
completely intermixed.} of the event any particular observed particle
has come.  If we {\em did} know from which side each particle had
come, we could use \MTTWO\ \cite{pubstransversemass,Barr:2003rg} to
place a lower bound on the mass of the two hypothesised outgoing
primary particles.  For a given event, even though we do not know the
``correct'' side assignments, there is nothing to prevent us trying
{\em all possible side assignments}, evaluating \MTTWO\ for each of
them, and then reporting the lowest value of \MTTWO\ so obtained.
This is in fact how \MTGEN\ is defined.

\subsection{Definition of \MTGEN}

\MTGEN\ is defined to be the smallest value of \MTTWO\
\cite{pubstransversemass,Barr:2003rg} obtained over all possible
partitions of momenta in $F$ into two subsets $\alpha$ and $\beta$ --
each subset representing the decay products of a particular ``side''
of the event.  Recall that \MTTWO\ is itself defined in terms of ${\bf
p}_T^\alpha$ and $m_\alpha$ (respectively the transverse momentum and invariant
mass of one side of the event), $ {\bf p}_T^\beta$ and $m_\beta$
(respectively the transverse momentum and invariant mass of the other side of
the event), and $\chi$ (the mass of each of the unobserved particles
which are supposed to have been produced on each side of the event) as
follows:
\begin{eqnarray}
&\mttwosq&({\bf p}_T^\alpha, {\bf p}_T^\beta,  {\bf \pmiss}_T ,
  m_\alpha, m_\beta, \chi)
  \equiv \nonumber \\
 &  \equiv &{ \min_{\slashchar{{\bf q}}_T^{(1)} +
\slashchar{{\bf q}}_T^{(2)} = {\bf \pmiss}_T }} {\Bigl[ \max{ \Bigl\{
m_T^2({\bf p}_T^\alpha, \slashchar{{\bf q}}_T^{(1)}; m_\alpha, \chi) ,\
m_T^2({\bf p}_T^\beta,  \slashchar{{\bf q}}_T^{(2)}; m_\beta, \chi) \Bigr\} }
\Bigr]}\label{MT2:MT2DEF}
\end{eqnarray}
where
\begin{equation} 
m_T^2 ( {\bf p}^{\alpha}_T, {\bf p}_T^\ntlinoone; m_\alpha, \chi) \equiv { m_{\alpha}^2 + \chi^2 +
2 ( E_T^{\alpha} E_T^\ntlinoone - {\bf p}_T^{\alpha}\cdot{\bf p}_T^\ntlinoone ) }
\label{MT2:MTDEF}
\end{equation}
in which \begin{equation}E_T^{\alpha} = { \sqrt { ({\bf p}_T^{\alpha}) {^2} + m_\alpha^2 }}
\qquad \hbox{ and } \qquad {E_T^\ntlinoone} = { \sqrt{ ({\bf
p}_T^\ntlinoone){^2} + \chi^2 } }
\label{MT2:ETDEF}\end{equation}
and likewise for $\alpha \longleftrightarrow \beta$.  With the above
definition (in the case $\chi = m_\ntlinoone$), \MTTWO\ generates an
event-by-event lower bound on the mass of the particle whose decay
products made up either of the two sides of the event, under the
assumption that the event represents pair production
followed by decay to the visible particles and an unseen massive
particle on each side.  When evaluated at values of $\chi \ne
m_\ntlinoone$ the above properties are retained approximately (see
\cite{pubstransversemass,Barr:2003rg}).  There exist events which
allow this lower bound to saturate, and so (in the absence of
background) the upper endpoint of the \MTTWO\ distribution may be used
to determine the mass of the particle being pair produced.

We caution the reader to avoid the trap of mistakenly concluding that
\MTGEN, as defined above, is a function of purely transverse (and not
also longitudinal) momenta of the visible particles in $F$.  On the
contrary, the definition above makes use of the z-momentum of every
visible particle.  Although this z-momentum plays no part in forming
the transverse {\em momentum}, ${\bf p}_T^\alpha$ or ${\bf
p}_T^\beta$, of either side in any partition, it nonetheless can play
a significant role in forming each side's {\em invariant mass}:
$m_\alpha$ or $m_\beta$.  (This caution is not unique to \MTGEN, but
is equally relevant for any situation in which \MTTWO\ is used where a
side consists of an ``effective'' particle composed of two or more
real particles.)  We note that it is possible to define a ``Truly
Transverse'' form of \MTGEN, which we shall denote ``\MTTGEN'' by
requiring, before evaluation begins, that each input four-momentum be
individually longitudinally boosted to a frame in which its
$z$-component of momentum is zero.  In effect this throws away each
particle's $z$-momentum, does not change its transverse momentum, and
reduces its energy so as to keep the particle's mass invariant.  It is
equivalent to evaluating \MTGEN\ in a ``transverse'' Minkowski space
with (1+2) dimensions rather than the usual (1+3) dimensions.

\section{Discussion}

\MTTWO\ has been used in the definition of \MTGEN, rather than a
function of the invariant masses of the two sides of the event, as
it is vital to account for the energy-momentum of unobserved
particles.  If an event has a non-zero total transverse momentum, then
either there were unobserved particles in the event, or some momentum
was mis-measured (or lost or created), or indeed a combination of
both.  \MTTWO\ takes these unobserved particles into account in hadron
colliders.  However, see the comment in the next section regarding use
in lepton colliders.

\subsection{Regarding specialisations}
There is room for some slight specialisation of \MTGEN\ to the
particular problem in hand.  \MTGEN\ need not always be calculated in
exactly the same way.

For example, if it were desirable to suppose that neither side of an
event could decay {\em entirely} into invisible particles, then one
might wish to impose the additional constraint that $F$ contain {\em
at least two} momenta, and that one should only consider partitions of
$F$ into {\em non-empty} subsets.  Events with fewer than two momenta
in $F$ could either be ignored or given \MTGEN\ values of
0.\footnote{When investigating particular classes of $R$-parity
violating supersymmetric models, an alternative specialisation might
be appropriate.  Here it might be desirable to suppose that none of
the decay products on either side of the event could go undetected (be
invisible).  In this case one might consider evaluating not \MTTWO\ in
the definition of \MTGEN\ but instead the larger of the two {\em
invariant masses} of each side of the event.  Note, however, that if
this were done, one would have to satisfactorily address the question
of where any observed missing transverse momentum had come from.
Under the supposition at work here, large missing transverse momenta
would indicate either large measurement error, or an event
incompatible with the assumptions, and might suggest that the event
should be disregarded or recalibrated.}

In an alternative specialisation, one could suggest constraining the
partitions of $F$ in such a way as to allow only those which meet
certain requirements in terms of conserved quantum numbers.  For
example, one might choose to reject assignments which make the
absolute value of the charges on either side of the event greater
than 1 or veto events where the total charge in $F$ is more than some
fixed value.\footnote{For example, at the LHC the charge of primary
interaction should be $-1$, $0$ or $+1$.}  One might try to restrict
on the basis of lepton number, arguing from the standpoint of lepton
universality (though this might be hard given the possibility of
unobserved neutrinos).

Finally, if \MTGEN\ were to be used at a lepton collider where the
momentum of the centre of mass was known to a reasonable precision, it
would be sensible to replace \MTTWO\ (which is a variable designed for
hadron colliders and so uses only transverse quantities in order to be
insensitive to longitudinal boosts) with a variable analogous to
\MTTWO\ but designed to make use of $z$-momenta.

\subsection{Avoiding over-specialisation}

Though one can tailor \MTGEN\ to the requirements or assumptions of a
particular investigation, it should be pointed out that the
``philosophy'' of \MTGEN\ is to avoid such specialisation where
unnecessary.

Most events that are actually of the pair-produced type we are
concentrating on will have a large number of partitions.  One and only
one of those partitions is correct, and it will produce an \MTTWO\
value that is appropriate (i.e.\ bounded above by the mass of the
initially produced particle).  All other (i.e.\ all wrong) partitions
will result in two or more particles from different ``sides'' of the
event being assigned to the same side.  These wrong partitions thus
{\em tend} to have have large \MTTWO\ values, if only for the reason
the union of
particles from opposite sides of the event tends to yield four-momenta
with very large invariant masses.  Most wrong partitions therefore
lead to \MTTWO\ values larger than that generated by the ``correct''
partition, and so excluding large numbers of these ``bad'' partitions
often has no effect on the eventual value of \MTGEN.

By avoiding complex specialisations, one can make the variable to a
large extent insensitive to a detector's ability (or inability) to
distinguish particle types.  This can make the variable useful for example
during early running when the detectors' abilities to determine particle
identification may not be well understood.

\subsection{Use as a cut variable}
\label{sec:cut}

Although \MTTWO\ was originally proposed as a variable for measuring
particle masses, it can also be used as a ``cut variable'' intended to
separate certain new-physics signals from Standard Model 
backgrounds.\footnote{When used in this way the parameter $\chi$ which represents 
the mass of the stable invisible particle is typically set to
zero. This is of course correct to a very good approximation for 
the neutrinos which produce the missing momentum in Standard Model events.}
This success is in part due an accidental conspiracy of three effects. (1)
\MTTWO\ tends to small values for back-to-back QCD-like events.  (2)
\MTTWO\ tends to small values when the missing transverse momentum is
small (which it is in much of QCD).  (3) \MTTWO\ tends to small values
when the missing momentum is parallel to one of the visible particles
fed to \MTTWO. This can easily happen in QCD when there are neutrinos
in a jet, or a single jet is mismeasured through inadequate
containment in a detector or passage through a crack region. By
accident rather than by design, therefore, \MTTWO\ tends to shift
badly measured and Standard Model events away from large values, which
are where the endpoints of the distributions containing new physics
are expected to be found.

\MTGEN\ shares those features of \MTTWO, as it is always bounded above
by \MTTWO\ for the correct partition.  \MTGEN\ may therefore also be
expected to find a role as a ``cut variable'', distinguishing events
by their inherent mass scale, and able to focus on better-measured
events.



\subsection{Comparison with other mass-scale variables:  $\meff$}

In the past it has been suggested that a good starting point for the
determination of the mass scale of these new particles is the
``effective mass'' distribution
\cite{Hinchliffe:1997iu,usemeffPhysicsTDR}.  There are a number of
slightly different definitions of $\meff$ and the phrase ``mass
scale'' (a comprehensive list and comparisons between them may be
found in \cite{Tovey:2000wk}) but a typical definition of
$\meff$ would be
\begin{eqnarray}
\label{eq:meff}
\meff & = & \ptmiss + \sum_{i} {p_{T(i)}},
\end{eqnarray}
in which $\ptmiss$ is the magnitude of the event's missing transverse
momentum and where ${p_{T(i)}}$ is the magnitude of the transverse momentum of
the $i$-th hardest jet or lepton in the event.

\par

All definitions of $\meff$ are motivated by the fact that new
TeV-scale massive particles are likely to be produced near threshold,
and so by attempting to sum up the visible energy in each event, one
can hope to obtain an estimate of the energy required to form the two
such particles.  Broadly speaking, the peak in the $\meff$
distribution is regarded as the mass-scale estimator.  

\subsubsection{Problems with $\meff$}
\label{sec:meff}
Although the effective mass is a useful variable, and simple to
compute, it has a few undesirable properties:

  $\meff$ can be sensitive to the beam energy and the proton's parton
  distribution functions (pdfs).  This is primarily because the
  desired correlation between $\meff$ and the mass scale relies on the
  assumption that the particles are produced near threshold.  While it
  is true that the cross sections will usually {\em peak} at
  threshold, they can have significant tails extending to $\sqrt{\hat{s}}$
  values considerably beyond the threshold value.  As a result, the
  $\meff$ distribution is broad, having a width similar to its mean.
  This smearing means that it is very hard to make {\em precise}
  statements about the mass scale from $\meff$ alone.  In contrast,
  the endpoints of \MTGEN\ are ``precision measurements'' in the sense
  that, up to the statistical error, the experimental resolution,
  and the systematics from mis-assignments between $F$ and $G$,
  the endpoint of an \MTGEN\ edge has a direct interpretation as the
  mass of the underlying generated particles.

  The smearing due to pdfs described above means further that in the
  upper tails of the $\meff$ distribution one has little chance of
  distinguishing the different contributions arising from {\em
  distinct} pair-production processes.  There, the components of the
  $\meff$ distribution are expected to blur together.  With \MTGEN\
  it is possible, though not guaranteed, that one will see multiple
  edges or changes in gradient in the upper parts of the spectrum
  allowing the masses of more than one high mass particle to be
  determined with precision.  The extent to which this will work in
  practice is likely to depend on the extent to which the \MTGEN\
  distribution is itself smeared by mis-assignments between $F$ and
  $G$.  Secondary edges at lower values of \MTGEN\ may, if not
  obscured by Standard Model backgrounds, also carry precision
  information on the masses of lighter particles.

In fairness to $\meff$, one should point that \MTGEN\ has its own
weaknesses.  Most obviously, it is a number of orders of magnitude
more costly\footnote{Though see more positive view of this in
\mysecref{sec:evaluating}.} to compute than $\meff$, for what could
turn out to be little gain.  Secondly, due to its \MTTWO\ dependence,
\MTGEN\ has an explicit dependence on the hypothesised mass(es) $\chi$
of any invisible decay products.  This makes it more complicated to
handle, as it is should strictly be termed an {\em event function} (of
$\chi$) rather than an {\em event variable}.  It should be said in
defence of \MTGEN, however, that it is good that this dependence on
$\chi$ is explicit and plays a precise role in offsetting the \MTGEN\
endpoints to locations that have a clear physical interpretation.
Though the $\meff$ variable itself is simpler to compute, lacking
$\chi$ dependence, the need to consider $\chi$ cannot be escaped when
using $\meff$ to draw conclusions about mass scales
\cite{Tovey:2000wk}.  Thirdly and finally we note that \MTGEN, as
canonically defined, may be more sensitive to high rapidity ISR than
$\meff$ (see \mysecref{sec:examples}).  Having said this, the {\em
truly transverse} form, \MTTGEN, should be much more tolerant of ISR
while retaining the theoretical properties of \MTGEN.\footnote{Despite
having $\mttgen \le \mtgen$ on an event-by-event basis, the position
of the upper kinematic {\em endpoint} of the \MTTGEN\ distribution
should be the same as the endpoint for \MTGEN\ (although with reduced
statistics at the endpoint) provided that it is kinematically possible
for the decay products of each side to be produced with vanishing
relative rapidity.}

\subsection{Example \MTGEN\ distributions\label{sec:examples}}

\begin{figure}
 \begin{center}
  \includegraphics[angle=90,width=13cm]{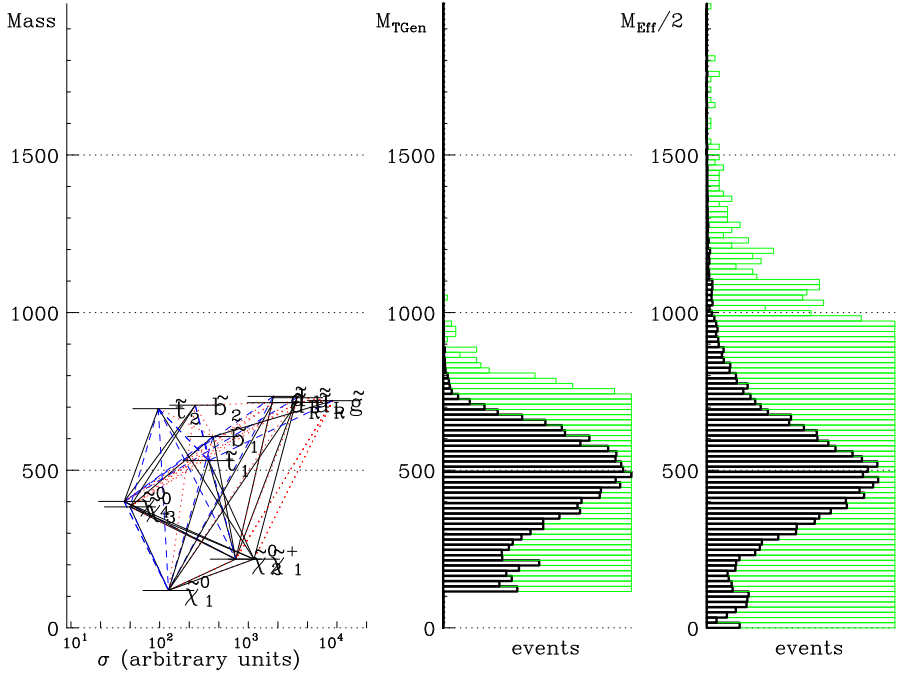}
\caption{
\label{fig:sps4_330_noisr}
On the left hand side is a graphical representation of the susy mass
spectrum of Snowmass point 4. The vertical positions 
of the particles indicate their masses.
The horizontal positions of the centres of the bars indicate the 
relative LHC production cross-section (arbitrary units).
The lines joining particles indicate decays with branching fractions
in the following ranges:
greater than $10^{-1}$ solid; 
$10^{-2} \rightarrow 10^{-1}$ dashed;
$10^{-3} \rightarrow 10^{-2}$ dotted.
The middle plot shows the distribution of our variable, \MTGEN, 
with \MTGEN\ increasing vertically to ease comparison with the spectrum.
The right hand plot shows the distribution of another variable, 
$\meff/2$, where $\meff$ is defined in \eqref{eq:meff}.
In both the \MTGEN\ and the $\meff$ plots, the lighter shading
shows the histograms with the number of events
multiplied by a factor of twenty, so that the detail in the upper tail may be seen.
}
\end{center}
\end{figure}

\begin{figure}
 \begin{center}
  \includegraphics[angle=90, width=13cm]{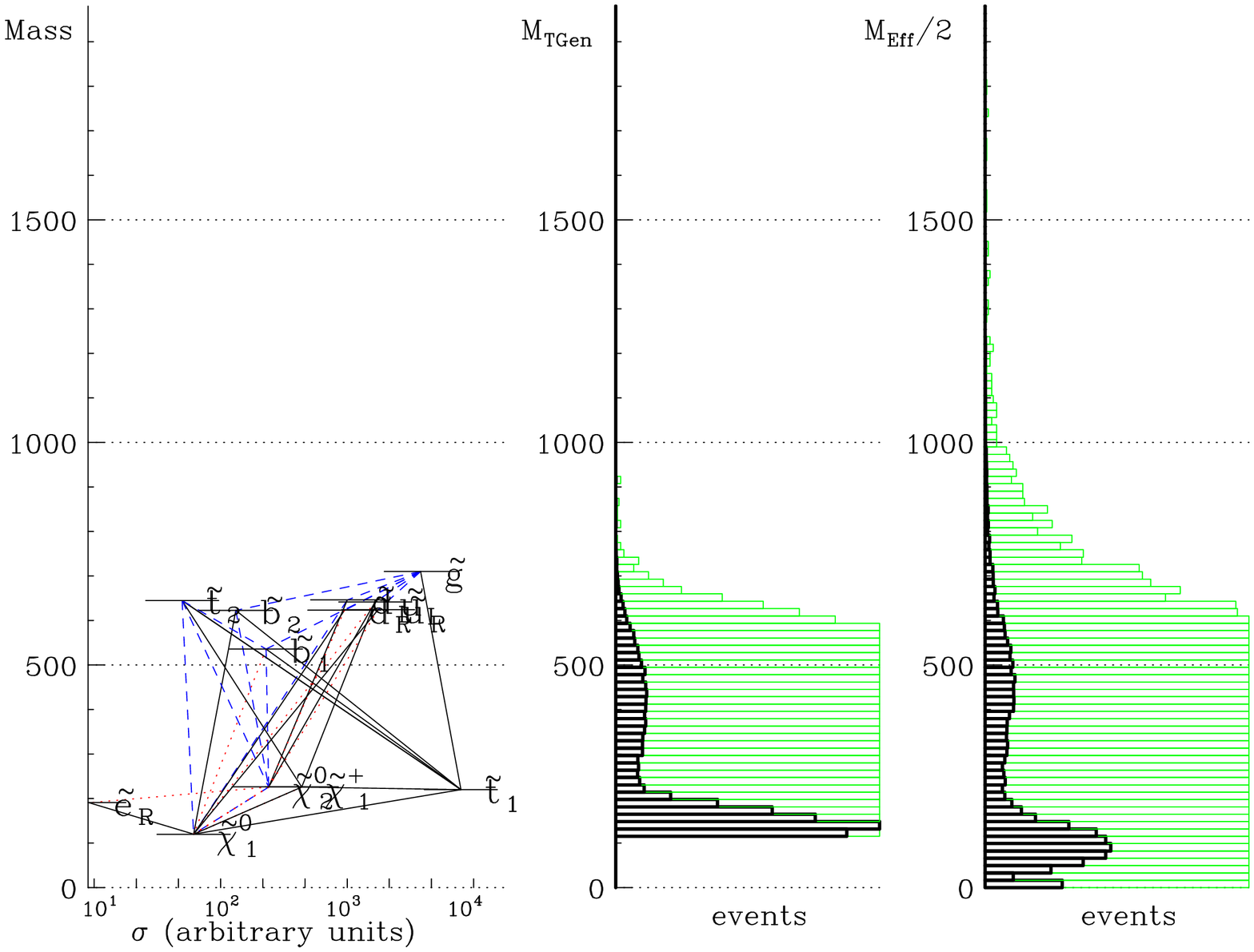}
\caption{
\label{fig:sps5_330_noisr}
As for figure \ref{fig:sps4_330_noisr}, but for Snowmass point 5.
}
\end{center}
\end{figure}

Simulations have been performed for several different supersymmetric
particle spectra, including the Snowmass points\cite{Allanach:2002nj},
for proton-proton collisions at LHC centre-of-mass energy of
$\sqrt{s}=$14~TeV.  The {\tt
HERWIG}\cite{Corcella:2002jc,Moretti:2002eu,Marchesini:1991ch} Monte
Carlo generator was used to produce inclusive unweighted
supersymmetric particle pair production events.  Final state particles
(other than the invisible neutrinos and neutralinos) were then
clustered into jets by the longitudinally invariant $k_T$ clustering
algorithm for hadron-hadron collisions\cite{Catani:1993hr} used in the
inclusive mode with $R=1.0$ \cite{Ellis:1993tq}.  Those resultant jets
which had both pseudo-rapidity ($\eta=-\ln\tan\theta/2$) satisfying
$|\eta|<2$ and transverse momentum greater than 10~GeV/c were used to
calculate \MTGEN\ and $\meff$.

\begin{figure}
 \begin{center}
  \includegraphics[angle=90, width=13cm]{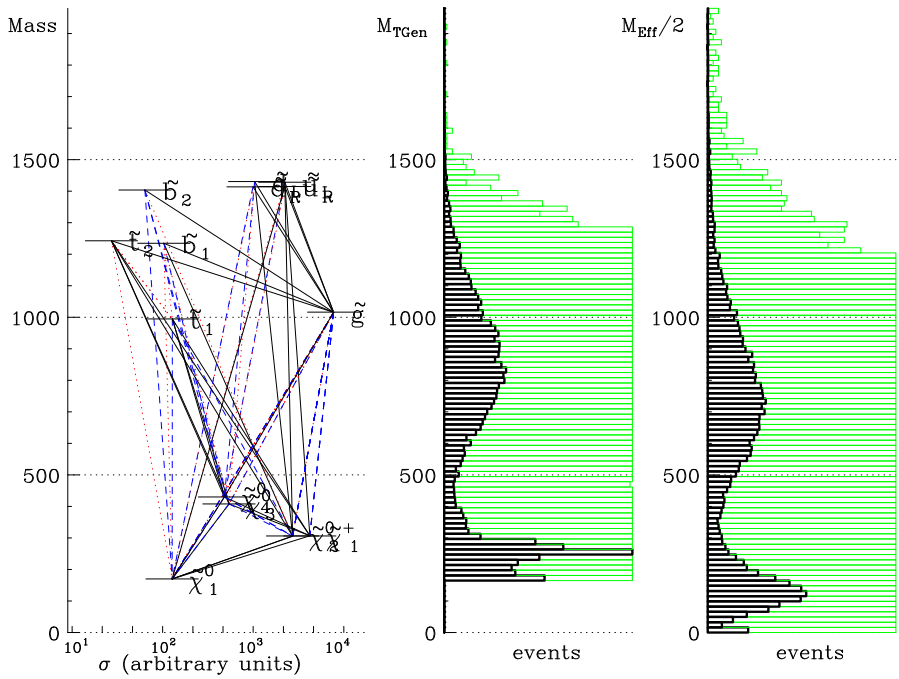}
\caption{
\label{fig:hwin0_330_noisr}
As for \myfigref{fig:sps4_330_noisr} but for a (non-Snowmass) 
point with a heavier sparticle spectrum, defined by
the mSUGRA parameters: 
\{$m_0=1200~$GeV, 
$m_\half=420~$GeV,
$\tan\beta=10$, 
$m_t=174~$GeV,
$\mu<0$\}
and with a spectrum generated using {\tt Isajet}\cite{Baer:1999sp} version~7.58.
}
\end{center}
\end{figure}

In figures \ref{fig:sps4_330_noisr}, \ref{fig:sps5_330_noisr}, and 
\ref{fig:hwin0_330_noisr}
we show the distributions which would be obtained for several 
different spectra if it were possible to accurately assign all visible 
momenta to the correct category $F$ or
$G$ (i.e.\ ``interesting final state momenta'' versus ``initial state
radiation''). 
The {\tt HERWIG} initial state radiation and underlying event have been 
switched off, and the parameter $\chi$ which is required to calculate \MTGEN\
has been set to the mass of the lightest supersymmetric particle.
The missing transverse momentum has been calculated 
from the negative vector sum of the momenta of the fiducial jets 
for reasons of computational efficiency
as described in \mysecref{sec:eval}.

It can be seen that the upper edge of the distributions 
gives a very good indication of the mass of the heaviest pair-produced sparticle.
Distributions from a variety of different supersymmetric points
show similar behaviour. This means that the 
position of the upper edge of the \MTGEN\ distribution 
can be used to find out about the mass scale of any 
semi-invisibly decaying, heavy, pair-produced particles.

Note that we have deliberately not used any information about 
the identity of the observed particles, 
so we do not know from this plot alone whether the
particles produced were squarks, gluinos, or indeed something completely different.
But we do have a very good indication that there is a particle being
pair-produced, 
and subsequently decaying to a mixture of visible and invisible particles,
and we have a good information about the mass scale at which this particle 
(or these particles) may be found.

Furthermore, in all three 
cases a change in slope can be observed at lower masses due to 
significant pair production of lower-mass particles (chargino and/or neutralino
pairs for \myfigref{fig:sps4_330_noisr} and \ref{fig:hwin0_330_noisr} 
or stop pairs for \myfigref{fig:sps5_330_noisr}).
Therefore it is possible in principle to extract from this distribution
information at several different mass scales.

The plots also demonstrate some of the 
the undesirable properties of the variable $\meff$. 
There is, as has already been shown in \cite{Tovey:2000wk}, 
some correlation between $\meff$ and the mass scale of particles being produced. 
However $\meff$ has a considerable tail at higher values 
caused by production of sparticles above threshold.

\begin{figure}
 \begin{center}
  \begin{minipage}[b]{.24\linewidth}\begin{center}
  \includegraphics[angle=90,width=3.5cm]{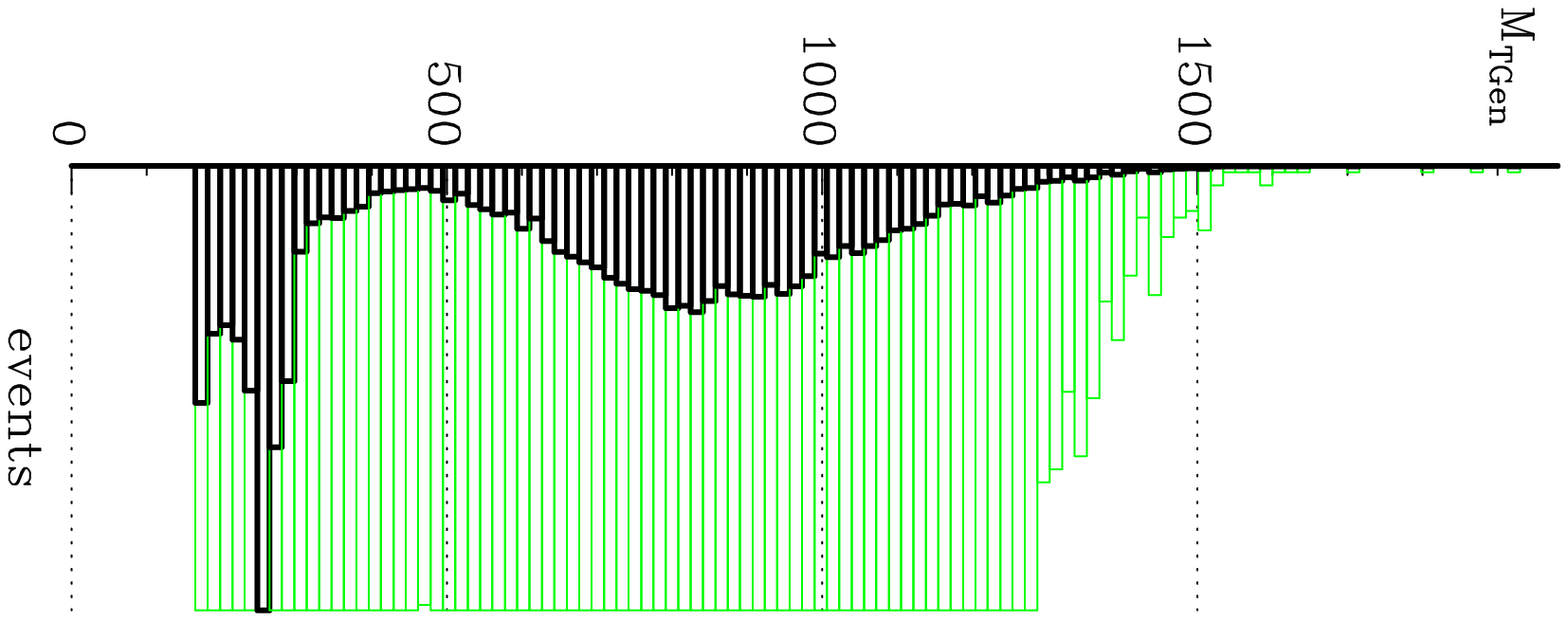} \\ {\bf (a)} 
  \end{center}\end{minipage}\hfill
  \begin{minipage}[b]{.24\linewidth}\begin{center}
  \includegraphics[angle=90, width=3.5cm]{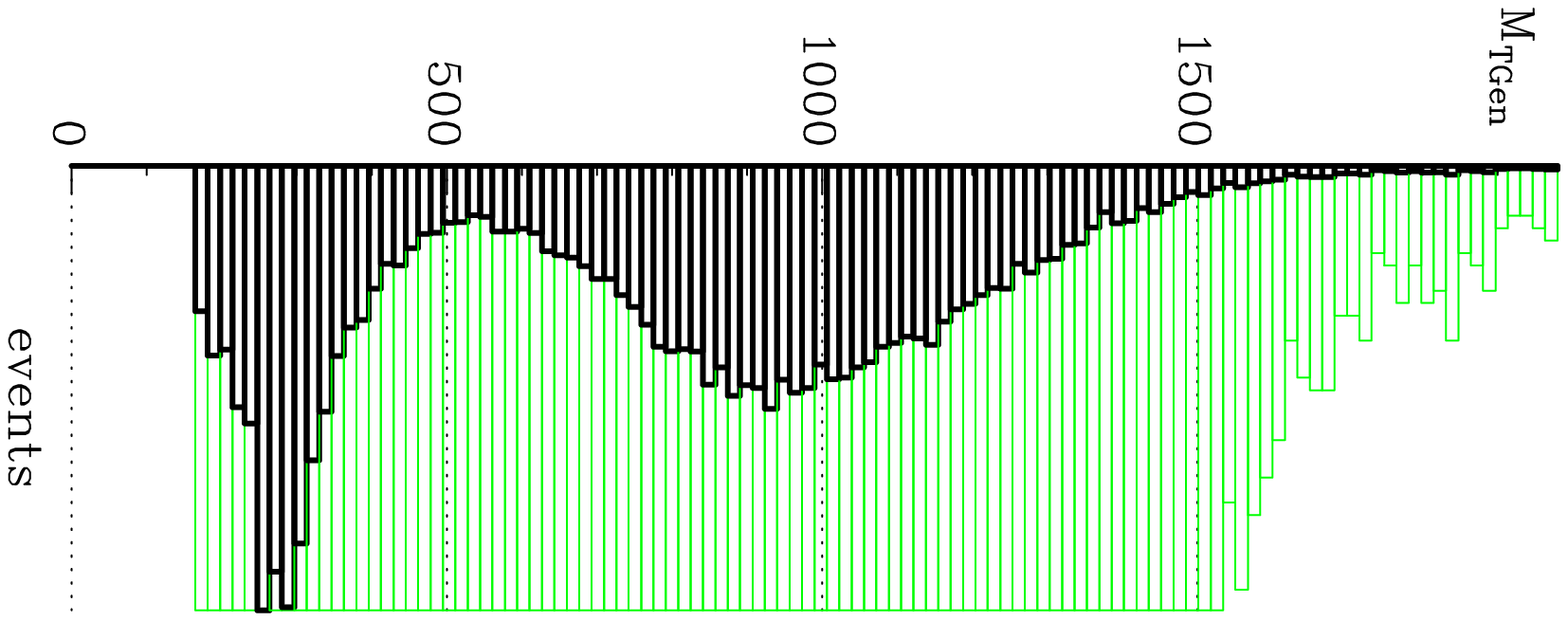} \\ {\bf (b)}  
  \end{center}\end{minipage}\hfill
  \begin{minipage}[b]{.24\linewidth}\begin{center}
  \includegraphics[angle=90, width=3.5cm]{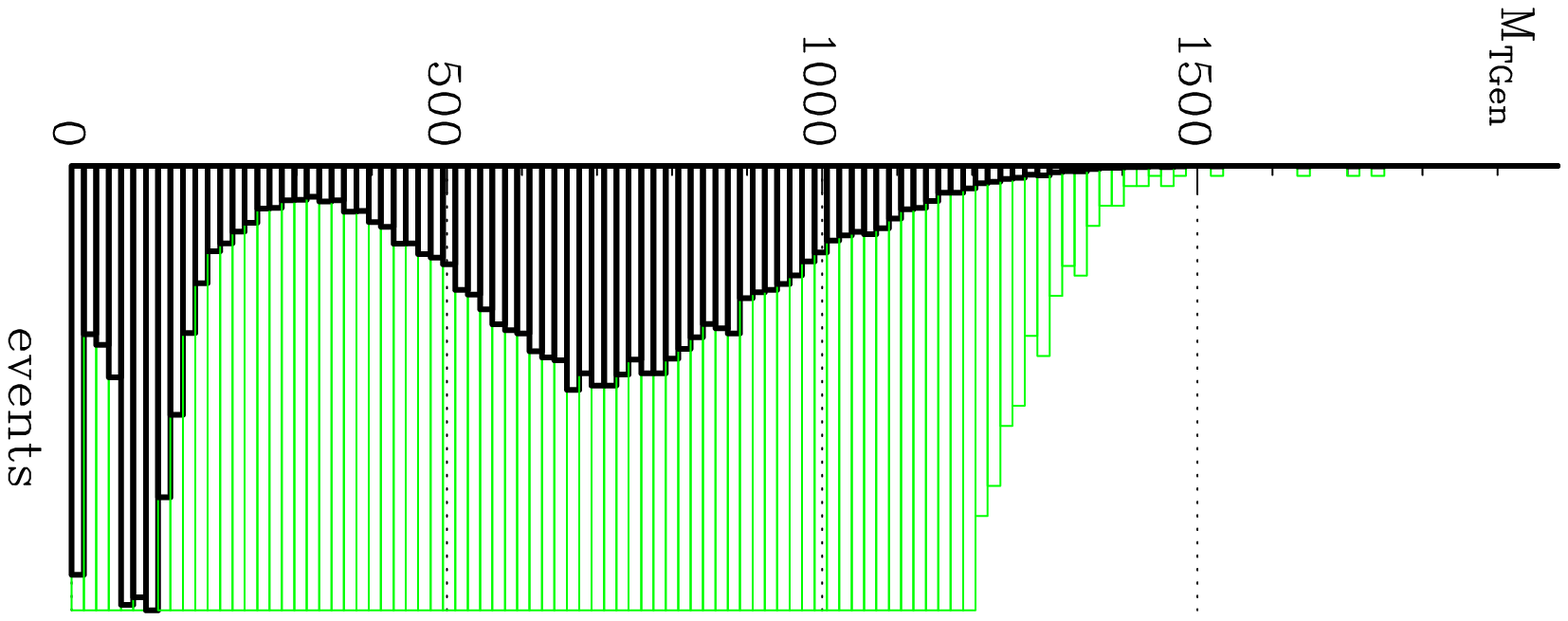} \\ {\bf (c)}
  \end{center}\end{minipage}\hfill
  \begin{minipage}[b]{.24\linewidth}\begin{center}
  \includegraphics[angle=90, width=3.5cm]{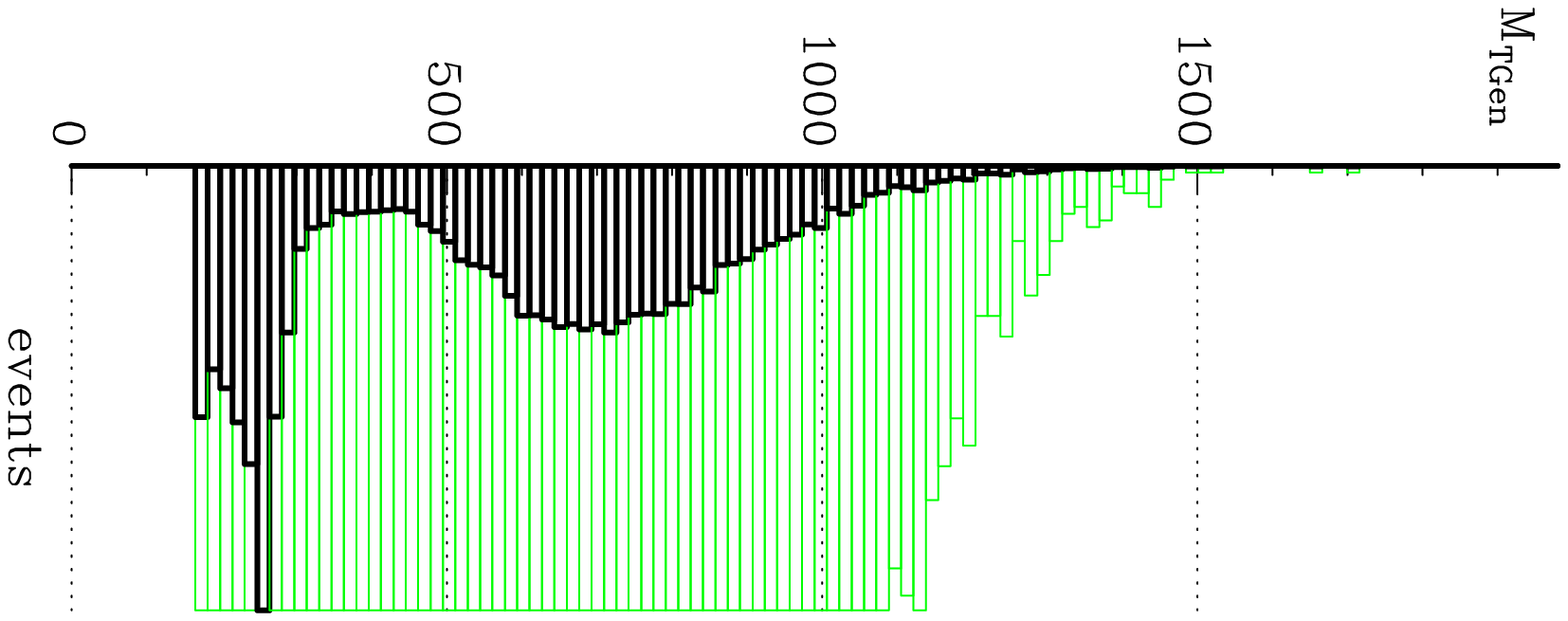} \\ {\bf (d)}
  \end{center}\end{minipage}\hfill
\caption{
\label{fig:hwin0_various}
\MTGEN\ distributions for the same spectrum as for figure
\ref{fig:hwin0_330_noisr}, but with different assumptions.
{\bf (a)} The idealised case, as described in text.
{\bf (b)} As for (a) but now with initial state radiation and 
the underlying event, and including particles from both categories $F$ and $G$
(``interesting'' and ``ISR/underlying event'') to form jets.
{\bf (c)} As for (a) but with the parameter, $\chi$, 
(corresponding to the mass of the invisible particle) set to zero.
{\bf (d)} As for (a) but using the sum of the transverse momenta of
the invisible particles for the missing transverse momentum.
}
\end{center}
\end{figure}
 
The effects of Standard Model processes (and
selection techniques required to reduce them) are beyond the scope of this paper. 
More detailed studies
using a complete set of Standard Model backgrounds and detailed 
detector simulation will be a important component of future work.
However we note that most SM events will have a small value of \MTGEN\ for the
reasons discussed in \mysecref{sec:cut}. 
Therefore, while these backgrounds might be expected to 
make it more difficult to extract information about lighter particles,
we do not expect them to significantly affect the upper edge
of the \MTGEN\ spectrum which contains the information about 
the heavier (here squark and/or gluino) particle masses.

The extent to which the approximations and assumption used in the
calculation of \MTGEN\ can be justified is explored in
\myfigref{fig:hwin0_various}.  If particles from initial state
radiation and the underlying event are allowed to ``pollute'' the
final state (\myfigref{fig:hwin0_various}b) there is some smearing of
the edges.\footnote{In the previous ``unpolluted'' plots, the
assignment of final state particles to class $F$ or $G$ was determined
from the kinematic history in the Monte Carlo event record.}  The
effect of adding the ISR and underlying event is to shift the apparent
position of the \MTGEN\ edge up by approximately 15-25\% for our point
(\myfigref{fig:hwin0_various}b).  This value represents an upper limit
to the uncertainty originating from these effects, which would be
obtained in the unlikely case where the corrections from ISR and the
underlying event were completely unknown.  The extent to which this
effect would be seen in experimental distributions will depend on the
details of the event selection (for example on the rapidity cut on the
jets).  We note also that, as anticipated, if the same plots are
generated for the ``Truly Transverse'' variant of the variable,
\MTTGEN, then the sensitivity of the endpoint to ISR is reduced (to
around 10-15\%) and there are proportionally fewer events at the
endpoint.  It remains to be seen whether the better strategy for the
future will be to invest time in improving ISR rejection while
focusing on \MTGEN, or to use variants like \MTTGEN\ with less
sensitivity to ISR but fewer statistics near the endpoint.

If the invisible particle mass is unknown, or if \MTGEN\ is being used as 
a selection variable, then the distribution with the mass parameter $\chi=0$
is most appropriate (\myfigref{fig:hwin0_various}c). 
In that case the lower limit of the distribution 
(which cannot drop below $\chi$) is pulled down toward the origin.
The change in the position of the upper edge is much smaller, 
as one would expect in the case where the total energy in the final state 
is dominated by visible particles. 

An efficient method of calculating \MTGEN\ 
which approximates the missing transverse momentum 
by the negative vector sum of the jet
transverse momenta 
(see \mysecref{sec:eval}) has been used for all these previous plots.
This can be justified by its good agreement with the corresponding distribution
obtained with the full numerical calculation of \MTGEN\
using the ``true'' missing momentum (\myfigref{fig:hwin0_various}d).

\subsection{Cross section constraints}

\MTGEN\ relies purely on the kinematics of four-momentum conservation
in each event.  It makes no use of cross section information, which
will therefore always remain a vital tool, orthogonal to \MTGEN, with
which to constrain the overall mass scale.

\subsection{Evaluating \MTGEN \label{sec:evaluating}}
\label{sec:eval}
Historically the main hurdle to the adoption of \MTGEN\ has been the
cost of evaluating it.\footnote{The authors first proposed \MTGEN\ for
the analysis of the ATLAS Blind Data Challenge which concluded with
the prize-giving at the ATLAS overview week in Prague, 2003, reported
in New Scientist. See ``Observations concerning the first ATLAS Blind
Data Challenge'' Barr, A J; Brochu, F M; Lester, C G; Palmer, M;
Sabetfakhri, A; Aug 2003} For each evaluation, \MTTWO\ typically
requires a numerical minimisation to be performed which can take from
a few thousandths to a few tenths of a second on most computers.  The
definition of \MTGEN\ requires this minimisation to be repeated up-to
$2^{(n_F)}$ times: once for each partition of $F$.  Since typical jet
definitions can lead to high jet multiplicities in supersymmetric
events, up to about 20 jets in some parts of parameter space, a naive
implementation can take many seconds, or even hours to calculate
\MTGEN\ for a single event.

For \MTGEN\ to become usable, it is important to find less time-consuming
ways in which the internal \MTTWO\ values can be calculated.  Ideally,
an analytic or closed form expression for \MTTWO\ is required to avoid
time-consuming numerical minimisations.

The authors were fortunate to be contacted in November 2006 by
Kyoungchul Kong and Konstantin Matchev (KKKM) \cite{KKKM}.  KKKM
informed the authors that they had derived an (undisclosed) analytic
expression for \MTTWO\ valid for the special case in which the missing
transverse momentum of the event was entirely balanced by the
transverse momentum of the two ``key visible particles'' input to
\MTTWO.  In other words, this special case corresponds to having no
net initial state radiation, or in the language of this paper $\left|
{ { \left( { \sum_{i=1}^{(n_G)} g_i^\mu} \right)}_T } \right| = 0$.
It was not until June 2007 that the authors realised that this happens
to be an interesting limit from the point of view of \MTGEN.  If all
reconstructed momenta are thrown into $F$, then $G$ is empty by
construction, and the special case is satisfied.  The more complicated
the event is, the more grounds one has for doing this, as the less
sure one can be as to the provenance of any individual particle.

It is important to confirm the existence of the analytic expression
claimed by KKKM and to perform timing tests using it in order to
support the claim that \MTGEN\ is now calculable in a reasonable time.
The authors were not able to obtain the full expression for \MTTWO\
from KKKM.\footnote{Though not releasing the full expression for
\MTTWO, KKKM did release the answer for the special-special case
where, in addition to $\left| { { \left( { \sum_{i=1}^{(n_G)} g_i^\mu}
\right)}_T } \right| = 0$ one also has the masses of the two visible
particles and the masses of the two hypothesised invisible particles
all equal to zero.  Though we did not make use of this information
when deriving our own expression (it cannot be used for \MTGEN\ as
most partition create event ``sides'' which have very large non-zero
masses) we are grateful to have been able to use it as a check.}  It
has therefore been necessary to re-derive (what is hopefully) the same
result independently in the Appendix.  The authors understand that
KKKM will release their own result in the near future
\cite{KKKMToAppear}.

Using the analytic form of \MTTWO\ derived in the Appendix, we find
that even with a naive ``try every partition of $F$'' algorithm we can
calculate \MTGEN\ for a 20-particle event in order one-second on a
typical personal computer.\footnote{This implementation is available
from the authors on request.}  The computation time scales as $2^N$
where N is the number of particles, so a 10-particle event can be
processed in one thousandth of that time. The authors find this is
more than fast enough to make \MTGEN\ as usable as other standard
event variables.\footnote{One would expect that, with more thought, it
would be possible to create faster algorithms.  Instead of trying all
possible partitions when minimising over \MTTWO\ values, it might be
possible to design an algorithm which flips momenta one-at-at-time
from one side of the event to the other -- hunting for the minimum.
More work would be necessary to determine whether such algorithms
could become stuck in local minima.}

\section{Conclusion}

In conclusion, we believe that \MTGEN\ could be an invaluable variable
for physicists working at the LHC, and other future colliders.  We
hope that if its usefulness is validated in subsequent dedicated
detector studies, it will become a standard tool in the Swiss army
knife for new physics searches.

\section{Acknowledgements}

The authors would like to thank Kyoungchul Kong and Konstantin Matchev
for giving us the confidence that there existed methods for
calculating \MTTWO\ in certain limits which were fast enough to allow
\MTGEN\ to be brought out of the cupboard and resurrected as a useful
event variable.  CGL would also like to thank Giacomo Polesello,
Tomasso Lari, James Frost, Martin White, Nic Barlow, Alan Phillips,
Sky French, Mario Serna and also members of the Cambridge
Supersymmetry Working Group for their thoughts and comments regarding
\MTGEN\ or drafts of this document.  The authors would also like to
thank the organisers of the 2007 Les Houches TeV Collider Workshop and
the 2007 ATLAS Overview Week in Glasgow for providing environments
supportive to the development of this work. AB wishes to thank the
Science and Technology Facilities Council (STFC) of the United Kingdom
for the financial support of his fellowship.

\appendix

\section{Appendix}

Here we derive an expression for $\mtTwo$ for the special case in
which the missing transverse momentum is {\em entirely} balanced by
the two visible particles' transverse momentum -- {\it i.e.} there
must be no ISR.  The particles themselves (and the hypothesised
missing particles) are allowed to have arbitrary masses.

Within this appendix we adopt the same conventions and definitions of
\cite{Barr:2003rg}.  In the language of 
that paper, the assumptions of
our ``special case'' can be phrased as ``$\Sigma=\sigma$''.  We will
begin, however, in the general case ($\Sigma \ne \sigma$) and only
introduce the simplification of the special case when we can no longer
make progress without it.  Beginning with the general case, then, we
have the following notation:
\begin{eqnarray}
\alpha^\mu &:& \mbox{ Lorentz 1+2 momentum of key visible particle 1 }
\\ \beta ^\mu &:& \mbox{ Lorentz 1+2 momentum of key visible particle
2 } \\ g ^\mu &:& \mbox{ Lorentz 1+2 momentum of junk or ISR } \\ p
^\mu &:& \mbox{ Lorentz 1+2 momentum of invisible particle (mass
$\chi$) produced with particle 1 }\\ q ^\mu &:& \mbox{ Lorentz 1+2
momentum of invisible particle (mass $\chi$) produced with particle 2
}\\ \Lambda^\mu &:& \mbox{ Lorentz 1+2 momentum of unit a mass
particle which is stationary in the lab frame }\\ \roots &:& \mbox{
real parameter (the reduced centre-of-mass energy 
from eq.~18 of \cite{Barr:2003rg})}
\end{eqnarray}
which are related by 
\begin{eqnarray}
\alpha^\mu + \beta^\mu + g^\mu + p^\mu + q^\mu = \roots
\Lambda^\mu.\label{eq:momcons}
\end{eqnarray}
Note that there is a potential ambiguity here between {\em real}
momenta, {\em measured} momenta, and {\em hypothesised} momenta.  In
this document, the quantities which are directly visible ($\alpha^\mu$,
$\beta^\mu$ and $g^\mu$) are taken to be be real momenta, or
equivalently to be measured quantities with zero measurement error.
Conversely $p^\mu$, $q^\mu$ and $\roots$ are quantities which cannot
be measured.  In this case these symbols refer to the {\em
hypothesised} neutralino momenta and/or {\em hypothesised} centre of
mass energies that are used throughout the process of describing the
event while attempting to calculate $\mtTwo$.

For simplicity, some derived quantities are also defined:
\begin{eqnarray}
\sigma^\mu = \alpha^\mu + \beta^\mu &:& \mbox{ Lorentz 1+2 momentum
sum of the two key visible particles } \\ \Delta^\mu = \alpha^\mu -
\beta^\mu &:& \mbox{ Lorentz 1+2 momentum difference of the two key
visible particles } \\ \Sigma^\mu = \sigma^\mu + g^\mu &:& \mbox{
Lorentz 1+2 momentum sum of everything seen in the detector }\\ B^\mu
= p^\mu + q^\mu &:& \mbox{ Lorentz 1+2 momentum sum of the two
invisible particles. }
\end{eqnarray}

We already know that the particular momenta of $p$ and $q$ which need
to be hypothesised to generate the value of $\mtTwo$ fall into one of
two categories.  Either they are in a ``balanced'' configuration in
which $(\alpha+p)^2 = (\beta+q)^2$ or the value of $\mtTwo$ is
achieved for an ``unbalanced'' configuration in which this is not
true.  It is easy to determine whether a given set of momenta
$\{\alpha^\mu, \beta^\mu, g^\mu\}$ generate $\mtTwo$ from a balanced
or an unbalanced configuration, and also easy to determine what
$\mtTwo$ is for the unbalanced cases.  We concentrate first,
therefore, on the harder case of how to calculate the value of
$\mtTwo$ if it has already been determined that it occurs in a
``balanced'' configuration.

\subsection{Balanced configurations}

In the ``balanced configuration'', the value of $\mtTwo$ will be the
minimum value of $(\alpha+p)^2$ over all allowed values of $\roots$
provided that the following constraints are satisfied:
\begin{eqnarray}
p^2 &=& \chi^2 \label{eq:one} \\ q^2 &=& \chi^2 \label{eq:two} \\
(\alpha+p)^2 &=& (\beta+q)^2 \label{eq:three}.
\end{eqnarray}
The first two constraints just put $p^\mu$ and $q^\mu$ on mass shell.
The final constraint is the one that makes the configuration mass
balanced.  When the constraint of equation~(\ref{eq:three}) is
satisfied we will refer to the both $(\alpha+p)^2$ and $(\beta+q)^2$
as $M^2$. The approach we will take will be to assume a fixed value of
$\roots$ and then solve the above three equations for $p^\mu$ and
$q^\mu$.  We then explicitly minimise the resulting value of $M$ by
varying $\roots$.  The resulting minimum value of $M$ is the value of
$\mtTwo$ we seek.

From equation~(\ref{eq:momcons}) we can see that for fixed $\roots$,
the value of $B^\mu$ is fully determined:
\begin{eqnarray}
B^\mu = \roots \Lambda^\mu - \Sigma^\mu \label{eq:part1}
\end{eqnarray}
and so the sum $p^\mu +
q^\mu$ is fixed.  We therefore choose to parametrise the three degrees
of freedom which $p^\mu$ and $q^\mu$ have collectively by writing them
in terms of an unknown Lorentz 1+2 vector $\gamma^\mu$ as follows:
\begin{eqnarray}
p^\mu = \half B^\mu + \gamma^\mu \label{eq:part5}\\
q^\mu = \half B^\mu - \gamma^\mu.\label{eq:part6}
\end{eqnarray}
Our stated intention of determining $p^\mu$ and $q^\mu$ for fixed
$\roots$ is therefore really a requirement to determine the three
components of $\gamma^\mu$, from the three constraints in
equations~(\ref{eq:one}), (\ref{eq:two}) and (\ref{eq:three}).  By
substituting the two equations above into equations~(\ref{eq:one}),
(\ref{eq:two}) and (\ref{eq:three}) it is easy to show that the
constraints on $\gamma^\mu$ are equivalent to the following:
\begin{eqnarray}
\gamma . B      &=& 0 \\
\gamma . \sigma &=& - \half \Delta . (B+\sigma) \\
\gamma^2 &=& - \quarter (B^2-4 \chi^2) .
\end{eqnarray}
The form of the above constraints motivates solving for $\gamma^\mu$ as
a linear combination of the three linearly independent vectors
$B^\mu$, $\sigma^\mu$ and $w^\mu = \epsilon^{\mu \nu \tau} \sigma_\nu
B_\tau$.  Doing this, one finds two possible solutions:
\begin{eqnarray}
\gamma^\mu = H^\mu \pm \what^\mu \sqrt{H^2 + \quarter(B^2-4 \chi^2) }
\label{eq:part2}
\end{eqnarray}
where
\begin{eqnarray}
\what^\mu = \frac{w^\mu}{\sqrt{-w^2}}  \label{eq:part3}
\end{eqnarray}
and
\begin{eqnarray}
H^\mu = {\frac{-\half \Delta.(B+\sigma)}{w^2} } {\left[ {(B^2)
\sigma^\mu - (\sigma.B) B^\mu} \right]} \label{eq:part4}.
\end{eqnarray}
Each solution corresponds to a kinematic configuration which is a
valid realisation of original mass constraints, but the value of
$M$ will almost certainly be different in each case.  Since our intention
is to find $\mtTwo$, we will eventually want to retain only the solution
which gives the smaller value of $M$.  As we do not yet know which
solution that is, we retain both for the moment.

We have now accomplished what we set out to achieve in step one.  For
fixed $\roots$ we have defined $B^\mu$ (with
equation~(\ref{eq:part1})).  This value may be substituted into
equations~(\ref{eq:part2}), (\ref{eq:part3}) and (\ref{eq:part4}) in
order to find $\gamma^\mu$.  In terms of $\gamma^\mu$ we can then find the
values of $p^\mu$ and $q^\mu$ (via equations~(\ref{eq:part5}) and
(\ref{eq:part6})) which lead to the so called ``balanced'' kinematic
structure in which both sides of the event have equal invariant mass
$M$.  All that now remains to do, is to {\em minimise} the value of
$M$ so-obtained over all allowed values of $\roots$.

It is at this stage that we now move the ``No ISR'' special case that
may be summarised as $\Sigma^\mu = \sigma^\mu$ or equivalently as
$g^\mu = 0$.  This change only affects terms with $B^\mu$'s in them as
these are the only quantities containing $\Sigma^\mu$.  Having made
the substitution $\Sigma^\mu \rightarrow \sigma^\mu$ there is a
substantial amount of cancellation within the expressions in terms of
which $\gamma^\mu$ is defined (equation~(\ref{eq:part2})).  The net
effect of this cancellation leaves $M$'s dependence on $\roots$ in the
relatively simple form:
\begin{eqnarray}
M^2 = E + A \roots \pm \lambda \sqrt{(\roots-D)^2-C^2}\label{eq:form}
\end{eqnarray}
for suitable values of the real quantities $A$, $C$, $D$, $E$ and
$\lambda$ which do not depend on $\roots$.  It is straightforward to
show that the minimum of this function occurs when $\roots$ takes the
value:\footnote{There is also a stationary point at $\roots = D - {C}/{\sqrt{1-\frac{\lambda^2}{A^2}}}$  but it can be shown that this is always unphysical.}
\begin{eqnarray}
\roots = D + \frac{C}{\sqrt{1-\frac{\lambda^2}{A^2}}}\label{eq:rootsatmin}
\end{eqnarray}
All that is needed to complete the evaluation of $\mtTwo$ in this
special case, then, is to determine the quantities $A$, $C$, $D$, and
$\lambda$.  ($E$ is not needed to calculate the value of $\roots$
which minimises $M^2$.\footnote{Added in 2009:  For completeness we
  note that $E = \chi^2 +
  \frac{1}{2}(m_\alpha^2+m_\beta^2)-\frac{1}{2}\sigma^2 +\half
\frac{(\Lambda.\Delta)}{|\sigpt|^2} \left[ \sigma^2 (\Lambda.\Delta) -
(\Lambda.\sigma)( \sigma.\Delta) \right]$.   We note further that once
the value of $\roots$ obtained in (\ref{eq:rootsatmin}) is
substituted into the expression for $M$ in (\ref{eq:form}) one finds
that $M^2_{\rm min}=E + A D + C \sqrt{A^2-\lambda^2}$.     })  We can then evaluate $\roots$ in terms of
these quantities, allowing in turn $B^\mu$, $w^\mu$, $H^\mu$,
$\gamma^\mu$, $p^\mu$ and finally $\mtTwo$ to be calculated. It may be
shown that the values needed are as follows:
\begin{eqnarray}
A&=& \half (\Lambda.\sigma) + \half
\frac{(\Lambda.\Delta)}{|\sigpt|^2} \left[ (\sigma.\Delta) -
(\Lambda.\sigma)( \Lambda.\Delta) \right], \\ C&=& \sqrt{|\sigpt|^2 +
\frac{\chi^2}{J}}, \\ D&=& \Lambda.\sigma \\ \mbox{and}\\
\lambda&=&\frac{\epsilon^{\mu \nu \tau} \Delta_\mu \sigma_\nu
\Lambda_\tau \sqrt{J}}{|\sigpt|}\\ \mbox{where} \\
J&=&\frac{|\sigpt|^2-(\Lambda.\Delta)^2}{4 |\sigpt|^2}
\end{eqnarray}
and the quantity $|\sigpt|^2$ is always evaluated in the lab
frame.\footnote{In other words $|\sigpt|^2 = (\sigma
. \Lambda)^2-\sigma^2$.}   Note that the condition expressed in
equation~(\ref{eq:three}) means that we can use a number of different
expressions to finally evaluate $\mtTwo$.  The simplest would be
$\mtTwo^2 = (\alpha+p)^2$ or $\mtTwo^2 = (\beta+q)^2$.  However it is
arguably nicer to preserve the explicit symmetry between the two sides
of the event by instead evaluating $\mtTwo^2$ as the average of these
two identical quantities.  If this is done, one ends up with
\begin{eqnarray}
\mtTwo^2 &=& \half (\alpha+p)^2 + \half (\beta+q)^2 \\ &=& \chi^2 +
\half(m_\alpha^2 + m_\beta^2) + \half(\sigma.B) + (\Delta.\gamma).
\end{eqnarray}
Note that it was the last of these forms which was used to generate
the statement of equation~(\ref{eq:form}).
\subsection{Unbalanced solutions}
As discussed in \cite{Barr:2003rg}, the value of $\mtTwo$ does not always arise from
a configuration of hypothesised momenta in which both sides of the
event have the same invariant mass.  These unbalanced solutions arise
if the momentum splitting which places one of the hypothesised
neutralinos at the same transverse velocity, ${\bf v}_t = {\bf p}_T/E_T$,
as its visible ``partner'' (thereby
minimising the invariant mass of that side of the event) causes
the invariant mass of the other side of the event (which is then fixed
by momentum conservation) to be even lower.  This statement is
generally true, and does not require the move to the $\Sigma=\sigma$
special case considered in the section dealing with ``balanced''
solutions.  Nevertheless, in order to write the full expression for
the $\Sigma=\sigma$ case we need to take these possibilities into
account.

\subsection{Putting all cases together}

We can now combine the two previous results into the following complete
expression for $\mtTwo$ valid for events in which the
missing transverse momentum exactly balances the transverse momentum
of the two important visible particles ({\it i.e.} valid for the case
$\Sigma=\sigma$ also known as ``no ISR'').
\begin{equation}
\label{eq:myfinalanswer}
\mtTwo^2 = \begin{cases}
(m_\alpha+\chi)^2 & \mbox{iff\qquad } (m_\alpha+\chi)^2 \ge (\beta + \tilde q)^2,  \\
(m_\beta+\chi)^2 & \mbox{iff\qquad } (m_\beta+\chi)^2 \ge (\alpha + \tilde p)^2,  \\
(\alpha+p)^2 & \mbox{or equivalently}\\
(\beta+q)^2 & \mbox{or equivalently}\\
\chi^2 +
\half(m_\alpha^2 + m_\beta^2) + \half(\sigma.B) + (\Delta.\gamma)& \mbox{otherwise}
\end{cases}
\end{equation}
where ${\tilde q}^\mu = (\sqrt{\chi^2+|{\bf \underline {\tilde q}}|^2}, {\bf \underline
{\tilde q}})$ with ${\bf \underline {\tilde q}} = -{\bf \underline \Sigma}-
\frac{\chi}{m_\alpha}{\bf \underline \alpha}$ 
and
${\tilde p}^\mu =
(\sqrt{\chi^2+|{\bf \underline {\tilde p}}|^2}, {\bf \underline {\tilde p}})$ with ${\bf
\underline {\tilde p}} = -{\bf \underline \Sigma}- \frac{\chi}{m_\beta}{\bf
\underline \beta}$ and in which $p^\mu$ and $q^\mu$ (which must not be
confused with the entirely different quantities ${\tilde p}^\mu$ and
${\tilde q}^\mu$ just mentioned!) are defined by
\begin{eqnarray}
p^\mu &=& \half B^\mu + \gamma^\mu, \\
q^\mu &=& \half B^\mu - \gamma^\mu, \\
\gamma^\mu &=& H^\mu \pm \what^\mu \sqrt{H^2 + \quarter(B^2-4 \chi^2) }
\end{eqnarray}
(choosing the sign which leads to the smaller value of $\mtTwo$) in which
\begin{eqnarray}
\what^\mu = \frac{w^\mu}{\sqrt{-w^2}}  
\end{eqnarray}
with
\begin{eqnarray}
w^\mu = \epsilon^{\mu \nu \tau} \sigma_\nu
B_\tau
\end{eqnarray}
and
\begin{eqnarray}
H^\mu = {\frac{-\half \Delta.(B+\sigma)}{w^2} } {\left[ {(B^2)
\sigma^\mu - (\sigma.B) B^\mu} \right]} .
\end{eqnarray}
and where we have taken in the ``no ISR no junk'' case
\begin{eqnarray}
B^\mu = \roots\Lambda^\mu - \sigma^\mu
\end{eqnarray}
having set
\begin{eqnarray}
\roots = D + \frac{C}{\sqrt{1-\frac{\lambda^2}{A^2}}}
\end{eqnarray}
where
\begin{eqnarray}
A&=& \half (\Lambda.\sigma) + \half
\frac{(\Lambda.\Delta)}{|\sigpt|^2} \left[ (\sigma.\Delta) -
(\Lambda.\sigma)( \Lambda.\Delta) \right], \\ C&=& \sqrt{|\sigpt|^2 +
\frac{\chi^2}{J}}, \\ D&=& \Lambda.\sigma \mbox{\qquad\qquad\qquad\qquad\qquad and} \\
\lambda&=&\frac{\epsilon^{\mu \nu \tau} \Delta_\mu \sigma_\nu
\Lambda_\tau \sqrt{J}}{|\sigpt|}
\end{eqnarray}
in which
\begin{eqnarray}
J&=&\frac{|\sigpt|^2-(\Lambda.\Delta)^2}{4 |\sigpt|^2}
\end{eqnarray}
and where the quantity $|\sigpt|=\sqrt{(\sigma . \Lambda)^2-\sigma^2}$ is
the magnitude of the visible transverse momentum evaluated in the lab frame.

\subsection{Addendum in light of arXiv:0711.4526 \cite{Cho:2007dh}}

(This section was added in 2009.  The purpose of the addition is to
draw readers' attention to a nice result of relevance to this paper, published in a later paper by
different authors \cite{Cho:2007dh}.  It is hoped that by placing the
forward reference to \cite{Cho:2007dh} here readers may be more likely
to use of the simplification that the result affords.) \par It was observed in  \cite{Cho:2007dh} that, after some simplification,
equation~(\ref{eq:myfinalanswer}) may be written in the following
equivalent but shorter form (still valid only for the case
$\Sigma=\sigma$ also known as ``no ISR'') :
\begin{equation}
\label{eq:asimpleranswer}
\mtTwo^2 = \begin{cases}
(m_\alpha+\chi)^2 & \mbox{iff\qquad } (m_\alpha+\chi)^2 \ge (\beta + \tilde q)^2,  \\
(m_\beta+\chi)^2 & \mbox{iff\qquad } (m_\beta+\chi)^2 \ge (\alpha + \tilde p)^2,  \\
\chi^2 + A_T + \sqrt{(1+\frac{4 \chi^2}{2 A_T
    -m_\alpha^2-m_\beta^2})(A_T^2-m_\alpha^2 m_\beta^2)} & \mbox{otherwise}
\end{cases}
\end{equation}
where the new quantity $A_T$ is defined by $A_T = \sqrt{m_\alpha^2 + {\bf \underline \alpha}^2}
\sqrt{m_\beta^2 +{\bf \underline \beta}^2}+ {\bf \underline
  \alpha}.{\bf \underline \beta}$  and where all other quantities are defined
as previously for (\ref{eq:myfinalanswer}).
It will be noted that $A_T$ is very closely related to the
contransverse mass of \cite{Tovey:2008ui}.  We note that it is proved
in \cite{Cho:2007dh} that the whole of (\ref{eq:asimpleranswer}) is
invariant under simultaneous equal magnitude but anti-parallel boost
of $\alpha$ and $\beta$ in the transverse Lorentz 1+2 plane.

\bibliography{bib}

\end{document}